\documentclass[usenatbib]{mn2e}
\usepackage{epsfig}
\usepackage{amsmath}

\begin{document}

\title[The Galaxy Bi-Modality Evolution in the sSFR - M Plane]
{On the Evolution of the Bi-Modal Distribution of Galaxies in the Plane of
Specific Star Formation Rate versus Stellar Mass}
\author[B. C. Ciambur, G. Kauffmann, S. Wuyts]
  {B. C. ~Ciambur $^{1}$\thanks{Email: bogdancc@MPA-Garching.mpg.de}, G. ~Kauffmann $^{1}$, S. ~Wuyts $^{2}$\\
  $^{1}$ Max Planck Institute for Astrophysics, Karl Schwarzschild Str. 1
85748 Garching, Germany\\
  $^{2}$ Max Planck Institute for Extraterrestrial Physics , Giessenbach  Str. 1
85741 Garching, Germany}
\date{\today}
\maketitle

\begin{abstract}

We have compared the observed distribution of galaxies in the plane of
specific star formation rate versus stellar mass with the
predictions of the Garching semi-analytic model at redshifts 0, 1 and 2. 
The goal is to test whether the implementation of radio mode AGN feedback, which is responsible for terminating the formation of 
stars in high mass galaxies, provides an adequate match to current high-redshift observations.
The fraction of quenched galaxies as a function of stellar mass in the models is in good agreement
with data at z=0 and z=1. By z=2, there are too few quenched galaxies with low stellar masses
in the models.
At z=2, the population of galaxies with
no ongoing star formation is clearly separated from the `main sequence'
of star-forming galaxies in the data.  This is not found in the models,
because z=2 galaxies with stellar masses less than $\sim 10^{11} M_{\odot}$ are predicted 
to host black holes with relatively low masses (less than $10^8 M_{\odot}$).
The current implementation of radio mode feedback from such black holes 
reduces the cooling rates from the surrounding halo, but
does not generate sufficient energy to stop star formation entirely.
We suggest that the models may be brought into better agreement with the data
if black hole growth is triggered by disc instabilities in addition to major mergers, and
if feedback mechanisms associated with the formation of galactic bulges act to
quench star formation in galaxies.  

\end{abstract}


\section{Introduction}

Deep galaxy surveys have revealed that in star-forming galaxies, star formation rate is 
correlated with stellar mass, giving rise to a relation  referred to as the 
`main sequence of star-forming galaxies' (\cite{Daddi07}; \cite{Noeske07}). 
This relation displays time evolution -- galaxies of fixed mass form stars at progressively 
lower rates at later times. The scatter around the relation is quite small ($\sim $ 0.3 dex) and exhibits little redshift dependence. The picture which emerges from this is that the star formation histories of most star-forming  galaxies have been relatively quiescent. Longer periods of continuous star formation may be interspersed with short periods of elevated star formation (starbursts)  resulting from interactions or  mergers (\cite{Kauffmann06}). The bursts are thought to contribute to the scatter around the main sequence. 

In contrast, massive galaxies that are `red and dead' occupy a different, 
but well-defined region of the SFR - mass plane. These objects can be 
understood from the point of view of their histories: they are likely the remnants 
of major mergers between star forming galaxies. Such mergers destroy disks and produce
bulges, and lead to starbursts if the progenitors are  gas-rich. The merger
may drive  accretion onto a central supermassive black hole, thus triggering
an active galactic nucleus (AGN). The effect of the consumption of gas into stars
during the merger  and feedback from supernovae and AGN `quench' these 
massive systems,  shutting off their star formation and thus reddening their colours. 

In this paper, we compare the distribution of galaxies in the specific star formation 
rate - stellar mass plane  from three different  observational galaxy catalogues with 
predictions from the \cite{Guo11} semi-analytic model of galaxy formation and 
evolution. We do this for three redshift intervals  to analyse the evolution of 
this relation and to understand the key processes which lead to it. Although satellite 
galaxies in clusters may have their star formation inhibited due to the removal of their 
gas reservoir by ram-pressure stripping or tidal interactions with their  neighbours  
(called satellite quenching), this process pertains to a relatively small
fraction of the more massive galaxies. In the semi-analytic models, the main driver of 
the quenching of very massive galaxies is AGN `radio mode' feedback (\cite {Croton06}, hereafter C06). This mechanism injects energy into the gas surrounding the  galaxy, 
preventing it from cooling, condensing onto the galaxy and subsequently forming stars. 
The efficiency of this type of feedback is dependent on the amount of gas 
in a hot phase within the halo (this is higher for more massive haloes), 
as well as on the mass of the accreting black hole,
which in turn depends on the host galaxy's merger history. Because both haloes and black
holes form hierarchically through mergers in the models, it is therefore important 
to check that sufficient massive haloes and black holes have had time to grow 
and produce the observed proportion of massive, quiescent galaxies at 
high redshifts (e.g. \cite{McCarthy04}; \cite{Toft09}; \cite{Williams10}). Such a 
comparison can therefore be a useful tool to probe this aspect of the model.

\section{Data}

\subsection{Observational Datasets}

The galaxy samples in this work were drawn from several surveys 
and three redshift bins were considered. The shallowest sample, with galaxies 
in a redshift interval of 0.025 $<$ \textit{z} $<$ 0.05 and referred throughout 
as \textit{z} $\sim$ 0 , was drawn from data as used in \cite{Wang11} - 
a catalogue constructed by matching galaxies with stellar masses 
greater than $10^{10} M_{\odot}$ from the sixth data release (DR6) 
of the Sloan Digital Sky Survey (SDSS) (\cite{York00}) with the 
fourth data release of the GALEX survey (\cite{Martin05}). 

The deeper samples were based on those used by \cite{Wuyts11}, hereafter 
referred to as W11. They were drawn from the GOODS-S and UDS catalogues, 
as described in their paper. Each of these two catalogues contained galaxies  separated in two bins of redshift, namely 0.5 $<$ \textit{z} $<$ 1.5 
and 1.5 $<$ \textit{z} $<$ 2.5. They are referred throughout as 
\textit{z} $\sim$ 1 and \textit{z} $\sim$ 2, respectively. 
The UDS catalogue was based on data from several different surveys (Cosmic Assembly 
Near-infrared Deep Extragalactic Legacy Survey (CANDELS), \cite{Grogin11}; 
\cite{Koekemoer11}; Spitzer Extended Deep Survey (SEDS), PI: 
G. Fazio; Ultra Deep Survey (UDS),  PI: O. Almaini;  
SpUDS - a Spitzer public legacy program on the UDS, PI: J. Dunlop) 
and ground-based data, on an area of $\sim$ 200 $arcmin^{2}$, which constituted 
the UDS area covered by CANDELS (see Galametz et al. (in prep.) for details on the photometric catalogue and W11 for the derived stellar population properties). The GOODS-S catalogue was constructed 
in a similar way, on an area from the GOODS-South field (Great Observatories 
Origins Deep Survey, a 148 $arcmin^{2}$ area centred on the Chandra Deep Field 
South) covered by CANDELS-Wide and CANDELS-Deep.

Where possible, W11 estimated the star formation rates using combined ultraviolet and
mid-infrared or far-infrared photometry. 
For objects lacking an infrared detection, they used stellar population synthesis model
fits to the UV-to-8 $\mu$m spectral energy distributions to estimate parameters
such as  stellar masses and SFRs. We refer the reader to their paper 
for further details. W11 constructed SFR - stellar mass planes and studied the
variation of structural properties of galaxies across the plane, finding for example 
a clear separation between the two galaxy populations in terms of Sersic index. 

In our work, we make use of the W11 data with a cut in 
stellar mass at $M_{\star} = 10^{10} M_{\odot}$, 
above which the two deep samples are complete. The number of galaxies  in 
each of the redshift bins is shown in Table \ref{table:numbers}.

\begin{table}
\caption{Number of objects in each field. In the case of the GOODS-S and UDS, the final 
number of galaxies after the stellar mass cut is displayed along with the initial, total number.} 
\centering 
\begin{tabular}{c c c c} 
\hline\hline                        %inserts double horizontal lines
Redshift & SDSS & UDS & GOODS-S \\ [0.5ex] 
%heading
\hline                 
$\sim$0 & 10456 & - & -  \\ 
$\sim$1 & - & 975/8033 & 748/5467  \\
$\sim$2 & - & 1000/6649  & 554/3763 \\[1ex] 
\hline\hline
\end{tabular}
\label{table:numbers} % is used to refer this table in the text
\end{table}

\subsection{Semi-analytic model}

The Garching semi-analytic model simulates galaxies starting from the merger trees 
of dark matter haloes of the Millennium Run (\cite{Springel05}). The latter is a cosmological, 
N-body, dark matter only simulation with $2160^{3}$ particles in a box 500 Mpc/h 
in size. The cosmological parameters used were based on a combined analysis of 
the 2dFGRS (\cite{Colless01}) and the first-year release WMAP data (\cite{Spergel03}) 
and are as follows: $\Omega_{m}$ = 0.25, $\Omega_{b}$ = 0.045, $\Omega_{\Lambda}$ = 0.75,
 $\sigma_{8}$ = 0.9, \textit{n} = 1 and $H_{0}$ = 73 $km  s^{-1}  Mpc^{-1}$. 
The semi-analytic model populates the haloes with galaxies, which 
evolve according to the merger histories of their parent haloes as well as 
according to analytic `recipes' governing the baryonic physics. These include gas 
cooling, star formation, supernova and AGN feedback, chemical enrichment, 
and the effects of galaxy interactions on star formation and bulge formation. 

The version of the model used in this analysis is that of \cite{Guo11}. The
galaxy parameters used in this work were retrieved from the publicly available Millennium 
database (\cite{Lemson06}). Two outputs of this model are available on the database: 
one associated with the Millennium simulation (described above) and a second 
with Millennium II (\cite{Boylan09}), which is a high-resolution version of the 
former with the same cosmological parameters and particle number but within 
a simulation box 100 Mpc/h in size. Here we used the outputs resulting from the
 Millennium merger trees, due to the fact that this simulation contains
a larger number of massive galaxies.  In order to assure good resolution towards 
the lower mass end, a cut was made on the number of dark matter particles 
per halo, i.e. \textit{Np} $\ge$ 300. This did not affect any results for
galaxies with stellar masses greater than $10^{10} M_{\odot}$. Catalogues from three 
simulation snapshots closest in redshift to the three
observational datasets  were retrieved, corresponding to \textit{z}=0, 0.99 and 2.07. 

\section{Analysis and Results}

\subsection{sSFR - Mass planes}
\begin{figure*}
\begin{center}$
\begin{array}{cc}
\includegraphics[scale=0.37]{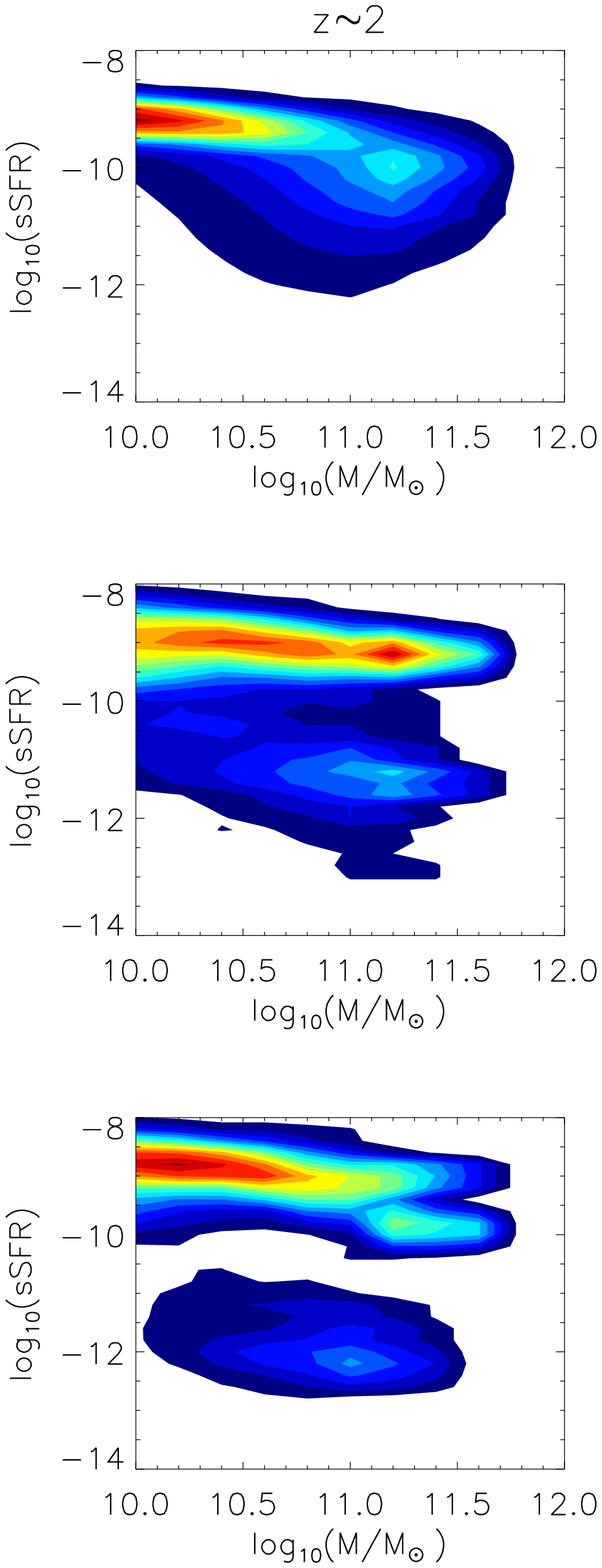} &
\includegraphics[scale=0.37]{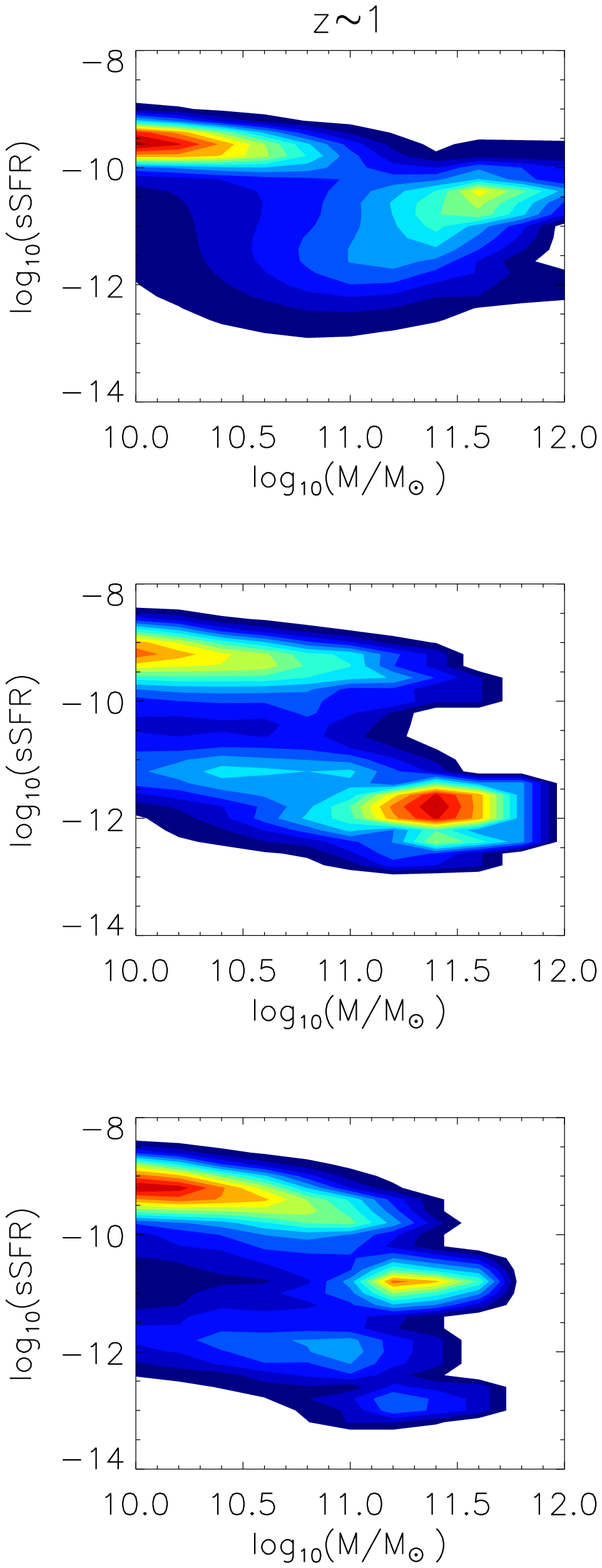} 
\includegraphics[scale=0.37]{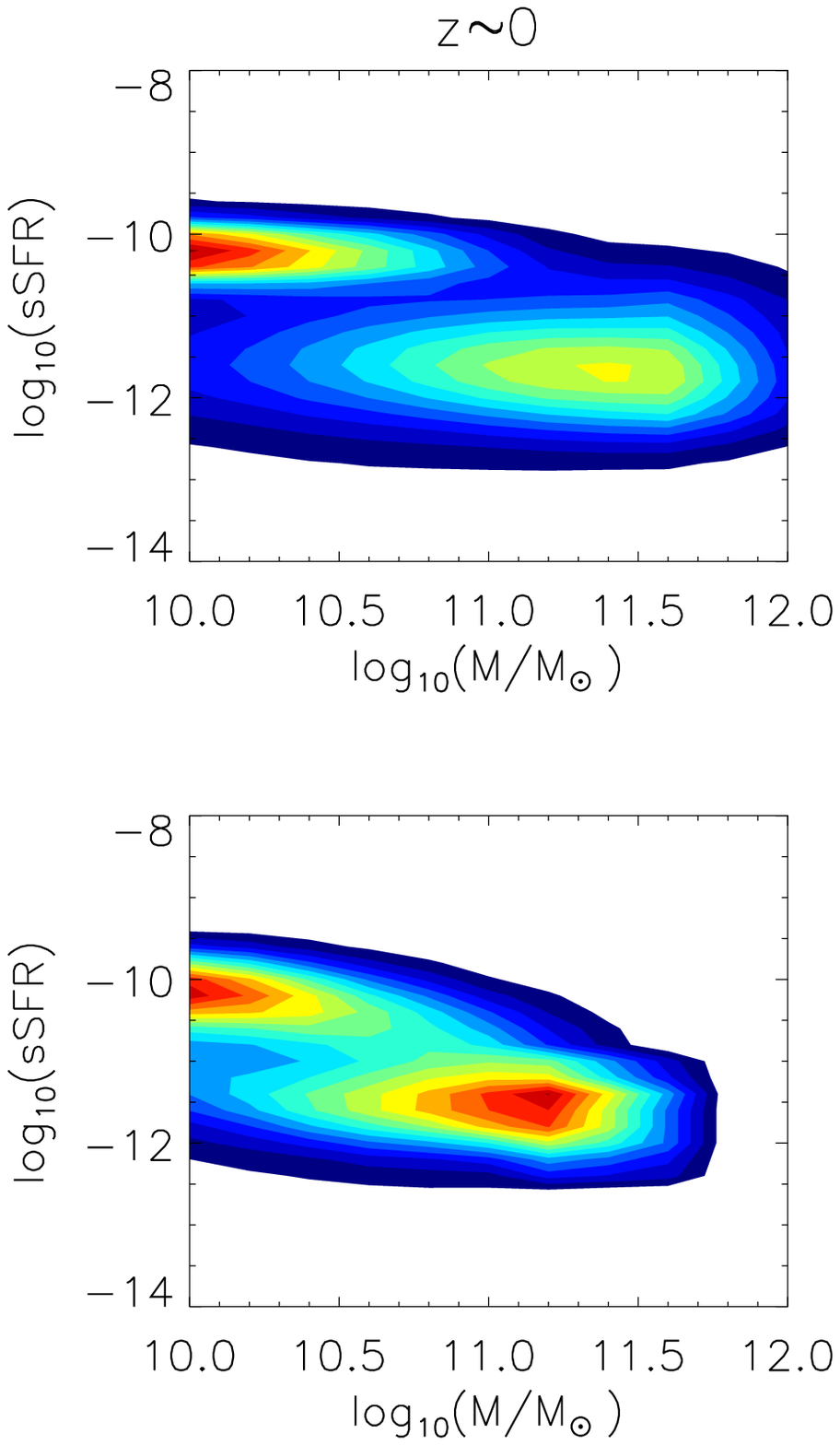} 
\end{array}$
\end{center}
\caption{Conditional density distributions of specific star formation rate (sSFR)
versus stellar mass,  as calculated in equation ~\ref{equ:normal}. Results are shown
for redshift 2 (left), redshift 1 (middle) and redshift 0 (right). 
For the two higher redshift bins the panels  are ordered, from top to bottom: 
Guo et al. model, UDS data, GOODS-S data. For the \textit{z} $\sim$ 0 panels on 
the right-hand side, the top plot is based on the model while the bottom on SDSS data.}
\label{fig:chpz012}
\end{figure*}

\begin{figure*}
\begin{center}$
\begin{array}{cc}
\includegraphics[scale=0.37]{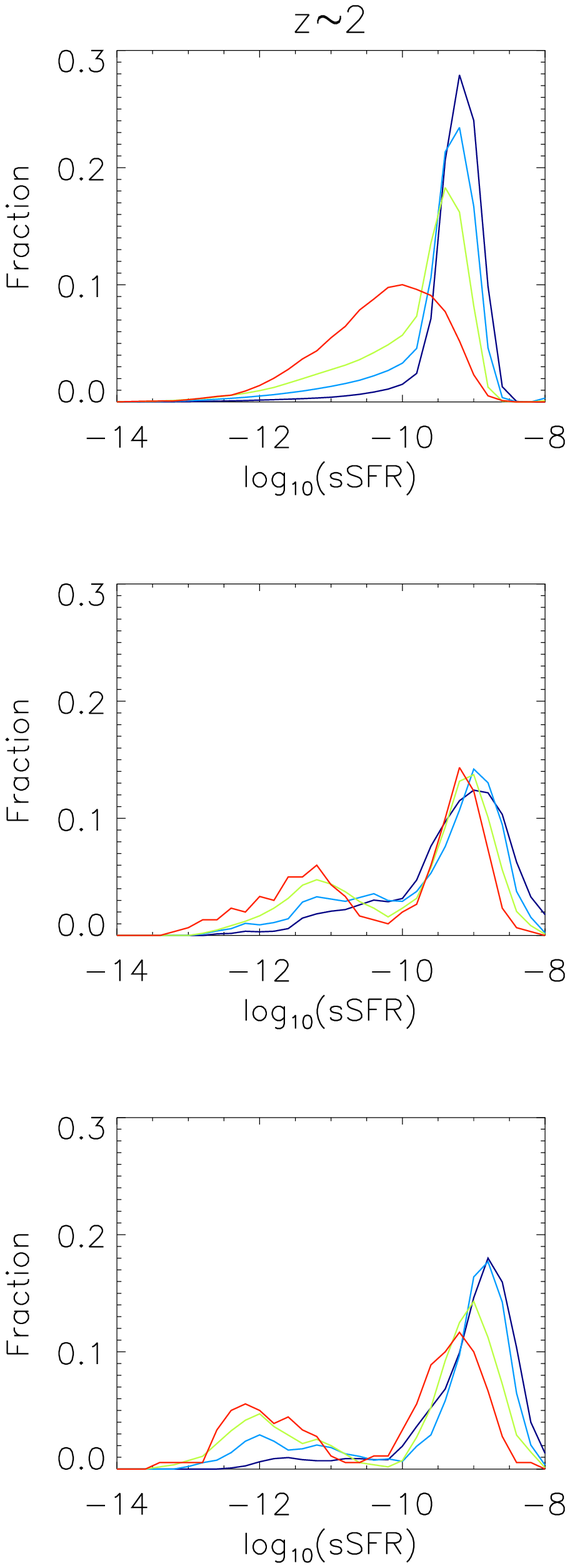} &
\includegraphics[scale=0.37]{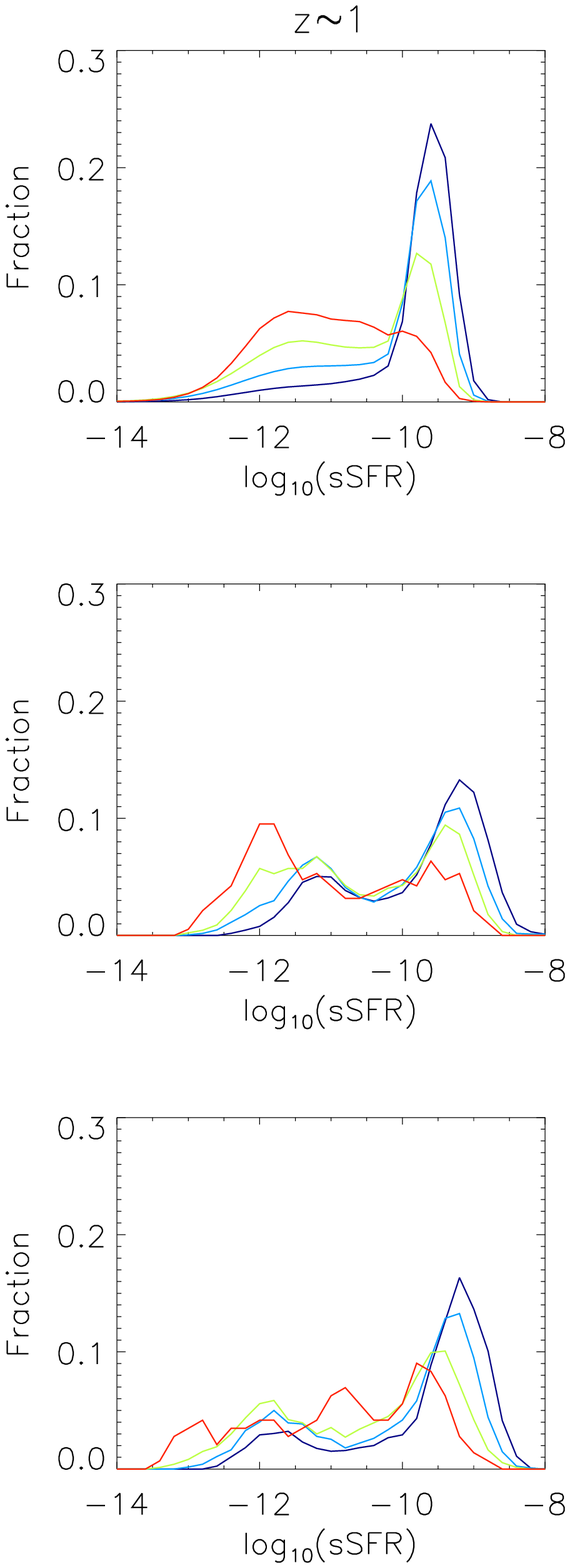} 
\includegraphics[scale=0.37]{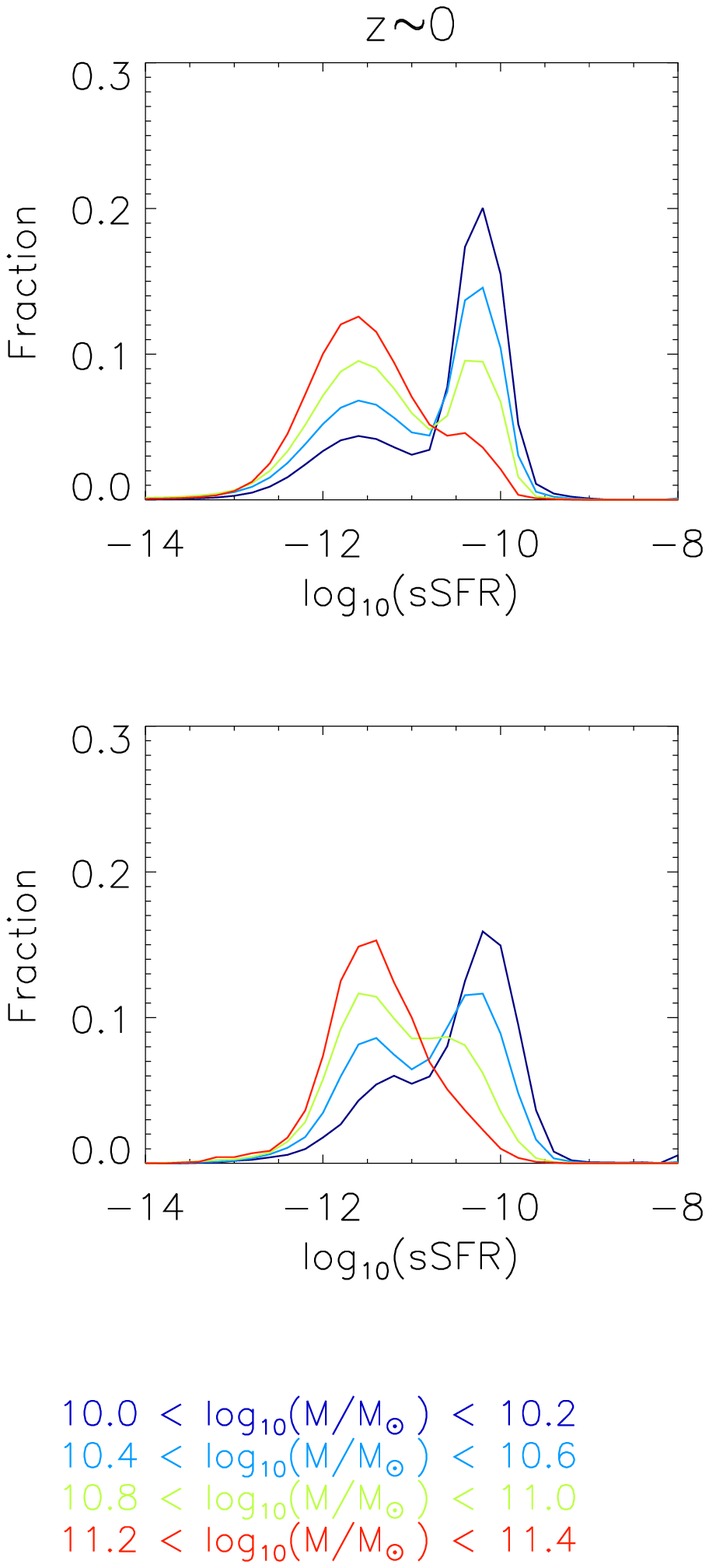} 
\end{array}$
\end{center}
\caption{Histograms of sSFR at fixed stellar mass. For clarity, only five out of the ten mass bins are plotted in each panel. As in Fig. \ref{fig:chpz012}, for the two higher redshift bins the panels  are ordered, from top to bottom: Guo et al. model, UDS data, GOODS-S data. For the \textit{z} $\sim$ 0 panels on the right-hand side, the top plot is based on the model while the bottom on SDSS data.}
\label{fig:hist012}
\end{figure*}

We first converted the log(SFR) - log($M_{\star}$) density plots in W11 into log(sSFR) - log($M_{\star}$)
density plots, i.e. star formation rate per unit mass or specific star formation rate 
(sSFR) as a function of stellar mass.  
The specific star formation rate is a measure of the time taken by a galaxy to form its stellar mass
at its current star formation rate. This quantity allows for a clearer
separation between the star-forming and quiescent populations, as will be seen later.

Next, all mass bins were normalised to unity, using the  equation 
\begin{equation}
  \label{equ:normal}
  N'(x,y)= \frac{N(x,y)}{\int \! N(x,y') \, \mathrm{d}y'}
\end{equation}
where $N(x,y)$ represents the count at position $(x,y)$, with $x$ $\equiv$ $M_{\star}$ and $y$ $\equiv$ sSFR. $N'$ is the normalised count, the quantity plotted in Fig. \ref{fig:chpz012}.\\
This allowed for an easier comparison between the data and the model, removing the need to scale  by the 
volume of the survey, which, in some cases was unknown. It also allows us to  visualize 
the bi-modality of the population on the density plots more easily (see \cite{Kauffmann03}). 
We note that  model galaxies with sSFR=0 were randomly re-assigned a value 
corresponding to the locus of the quenched population in the data. 
This was done in order to account for the fact that the observations always have a 
lower threshold in sSFR, set by the range of models used in the SED fitting. 
The effect of this was negligible at high redshift but more important at z $\sim$ 0, 
where AGN feedback completely halts star formation in a significant number of galaxies in the simulations. 

Fig. \ref{fig:chpz012} shows that the semi-analytic model is in qualitative agreement with the data
in that there is a clear  main sequence of star forming galaxies and another
separated population of massive, red and dead galaxies. A more detailed
comparison reveals that at high redshifts the main sequence appears to lie at lower sSFR 
compared with the data. This discrepancy is well-documented in the literature
(e.g. \cite{Damen09}).  Further, in the models the quenched peak appears to be located very close
to the main star forming sequence at early times and then travels down in time towards 
lower star formation rates. In the observations, the quenched galaxies are well 
separated from the star-forming ones even at \textit{z} $\sim$ 2.

\subsection{Quenched fractions as functions of stellar mass}

In order to quantify the evolution of the quenched population with redshift, 
the sSFR-M$_{\star}$  plane was divided into slices of constant stellar mass and 
histograms of the sSFR distribution at fixed mass were constructed (Fig. \ref{fig:hist012}).
These histograms can easily be decomposed into two components: one peak 
corresponding to the star-forming galaxies, and the remainder corresponding to the 
quenched population.
The objective of this exercise
is to compute the fraction of the population in the quenched state, 
per stellar mass bin, and ascertain whether the model fraction matches 
the observations, and whether model and observations  evolve the same way
as a function of redshift. 

\begin{figure}
\begin{center}
\includegraphics[scale=0.4]{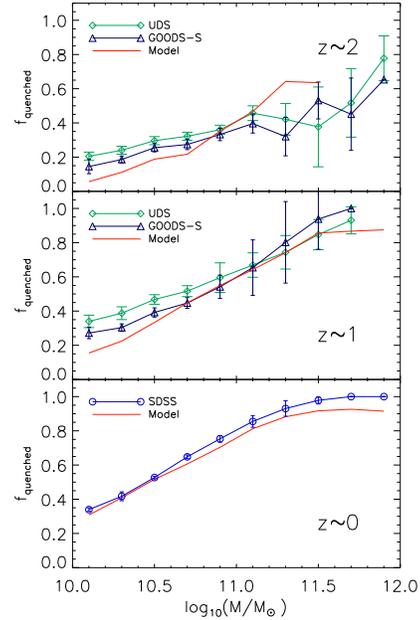} 
\caption{The fraction of quenched galaxies as a function of stellar mass. 
The red line with no symbols, which corresponds to the model, includes 
sSFR = 0 galaxies from the simulation in the quenched population. 
The error bars on the data points were computed using bootstrap resampling.}
\label{fig:fractions}
\end{center}
\end{figure}

A separation between the two peaks in sSFR was made visually and a Gaussian 
was fitted to the star-forming peak; the Gaussian was subtracted from the total 
distribution (which was by construction normalised to unity), 
leaving the remainder to be interpreted as the quenched fraction. 
Figure \ref{fig:fractions} shows the quenched fraction for both data and model at 
various stellar masses and redshifts. The error bars on the data were 
computed by bootstrap-resampling the original distribution (500 bootstrap subsets 
were used). Overall, the model agrees qualitatively with the observations. 
At redshift $\sim$ 0, the model slightly underpredicts the quenched fraction at 
high masses whereas at higher redshift, the model \textit{under}predicts 
the fraction at low masses and \textit{over}predicts it at high masses. This, taken in conjunction with the information in Figure \ref{fig:hist012}, leads to conclude that the bi-modality persists in the data over a broad mass range, whereas the model only tends to be bi-modal over a mass interval which narrows progressively with increasing redshift (at $ z \sim$ 2 the transition is quite sharp at $M_{\star} = 10^{11 - 11.2} M_{\odot}$). Outside of this interval, a single mode is preferred.

\subsection{Peak separation and its evolution}

\begin{figure}
\begin{center}
\leavevmode
\includegraphics[scale=0.4]{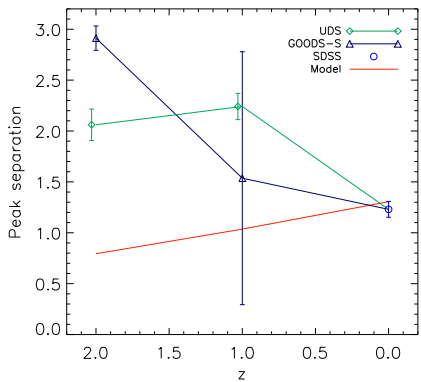}
\caption{The separation between the star-forming main sequence and the quenched peak 
in units of $\log$ sSFR, as a function of redshift. As in Fig. \ref{fig:fractions}, the error bars were calculated by bootstrap re-sampling.}
\label{fig:sep}
\end{center}
\end{figure}

Although the quenched fractions indicate a qualitative similarity between the 
model and the observations, they are not sensitive to the exact shapes of the 
distributions, or the separation between the star-forming main sequence  
and quiescent peaks, which is visibly different between model and data
in Figure 1 at higher redshifts. The evolution of the mean peak separation (averaged over all stellar masses) is shown in Fig. \ref{fig:sep}. 
It is noteworthy that the general trend in the data is opposite to that 
in the model. In the observations, the star-forming main sequence and the quenched 
population are well separated at high redshift and gradually 
approach each other with time. This approach is mainly due to the
decrease in star formation rate on the main sequence at lower
redshifts. In the model, there is a slight increase in the separation of the
peaks with time. This is caused by the fact that residual star formation
rates in the quenched population are larger at high redshifts than at low redshifts.

\section{Discussion and Conclusions}

We have compared the distribution of galaxies in the plane of
specific star formation rate versus stellar mass with the
predictions of the Garching semi-analytic model, for three redshift bins. 
Our  main findings are listed below.

\textit{i.)} There is qualitative agreement between models and data in that there is
a well-separated  main sequence of star-forming galaxies at low stellar masses
and a quenched population at higher masses at all three redshifts in the semi-analytic models.

\textit{ii.)} The fraction of quenched galaxies as a function of stellar mass in the models is in good agreement
with data at z=0. At higher redshifts, there are too few quenched galaxies with low stellar masses
in the models.

\textit{iii.)} The evolution in the separation between the star-forming and quenched populations is not            
reproduced by the models. In the data, the separation decreases towards low redshift, whereas
the the models it remains roughly constant. 

\begin{figure*}
\begin{center}
\leavevmode
\includegraphics[scale=0.4]{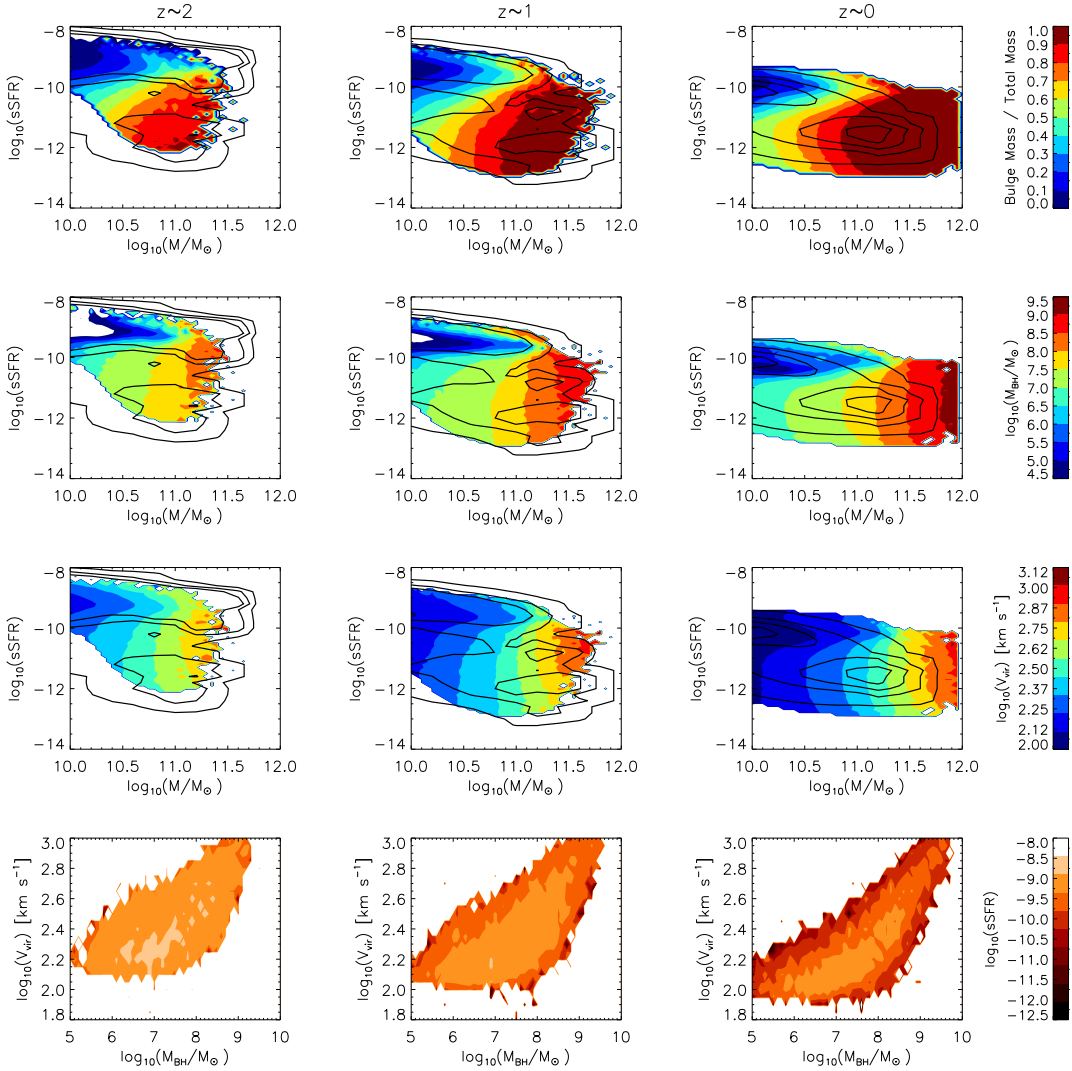}
\caption{Average galaxy properties across the sSFR - $M_{\star}$ planes, and the dependence of quenching on black hole mass and halo virial velocity. The upper three rows correspond to the sSFR - $M_{\star}$ planes  with conditional density of galaxies from the observational datasets (as in Fig. \ref{fig:chpz012}) overplotted as the black empty contours. The filled contours map the distribution of average bulge mass to total mass ratio (upper row), black hole mass (second row) and virial velocity (third row).  The bottom row shows the distribution of $M_\mathrm{BH}$ and $V_\mathrm{vir}$, colour-coded by sSFR. Every row has fixed contour levels to illustrate the redshift evolution of the quantities in question.}
\label{fig:mbh_vvir2}
\end{center}
\end{figure*}

Perusal of Figure 1 shows that in the observations, the quenched population gains
more and more members as time progresses, but it remains clearly separated from 
the star forming population even at early times. Moreover, it is clear that in the observations, the 
quenched population extends across the entire stellar mass range from
$10^{10} M_{\odot}$ to $\sim 2 \times 10^{11} M_{\odot}$ at all redshifts (even though the quenched fraction increases significantly towards higher masses). 
In contrast to this, in the models the quenched population becomes progressively more separated from
the star-forming population towards lower redshifts. In addition,  the quenched
population of high redshift galaxies in the models is confined to galaxies with higher stellar masses
than at low redshifts.

Let us now consider  the implementation of the AGN `radio-mode' feedback mechanism
in the models. This type of black hole activity results from accretion of hot gas 
from a static halo around the host galaxy, described in detail by C06:

\begin{equation}
  \label{equ:feedback1}
  \dot{M}_\mathrm{BH} = \kappa \,\frac{M_{BH}}{10^8 M_{\odot}h^{-1}}{\left(\frac{V_\mathrm{vir}}{200 km s^{-1}}\right)}^3 \frac{f_\mathrm{hot}}{0.1}
\end{equation}
where $\mathrm{M_{BH}}$ is the mass of the black hole, ${V_\mathrm{vir}}$ is the virial velocity of the parent dark matter halo (or subhalo) and \textit{$f_{hot}$} is the fraction of the total halo/subhalo mass in the form of hot gas. Here $\kappa$ is a free parameter which controls the efficiency of accretion, and it is taken to be 1.5 $\times 10^{-5} M_{\odot} yr^{-1}$. The black hole luminosity is given by: 

\begin{equation}
  \label{equ:feedback2}
  L_\mathrm{BH} = \eta \, \dot{M}_\mathrm{BH} \, c^2 
\end{equation}
where $\eta$ = 0.1 is the efficiency with which mass produces energy near the event horizon. The effect of this accretion is assumed to be injection of heat into the surrounding environment, preventing gas to cool, condense onto the galaxy and form stars. The effective cooling rate of gas is then affected in the following way:

\begin{equation}
  \label{equ:feedback3}
 \dot{M}_\mathrm{cool,eff} = \dot{M}_\mathrm{cool} - \frac{L_\mathrm{BH}}{\frac{1}{2}V_\mathrm{vir}^2}
\end{equation}

This feedback mechanism is assumed to be quiescent and continual and is most effective at late times and for high black hole masses. In equation \ref{equ:feedback1}, $f_{hot}$ has little effect on the accretion rate and is approximately constant for $V_{\mathrm{vir}}$ $\ge$ 150 $km \ s^{-1}$ at redshift 0, as shown in C06.
(We note that because this threshold in $V_{\mathrm{vir}}$ is actually at a fixed value of $M_{vir}$, it evolves as $(1+z)^{\frac{1}{2}}$) . 
This means that gas cooling suppression, and hence quenching, depends mostly 
on $M_{\mathrm{BH}}$ and $V_{\mathrm{vir}}$, which in turn depend on the history of the galaxy's parent halo. 

The top three rows of Fig. \ref{fig:mbh_vvir2} illustrate the average $M_{\mathrm{bulge}}/M_{\star}$ ratio, $M_{\mathrm{BH}}$ and 
$V_{\mathrm{vir}}$ for model galaxies as a function of
position in the sSFR - $M_{\star}$ plane. The bottom row represents the 
distribution of $M_{\mathrm{BH}}$ and $V_{\mathrm{vir}}$, colour-coded by sSFR. 
At a given redshift, the quiescent galaxies are more bulge-dominated, host more massive black holes\footnote{There is an exception to this if the galaxy in question has undergone a recent gas-rich merger. In this case there is also a starburst associated with the merger which boosts the star formation rate of the galaxy.}  and reside in haloes with larger virial velocities than the star-forming ones. The trends of the latter two parameters with redshift go in opposite directions: at a fixed location in the sSFR - $M_{\star}$ plane,  black hole masses grow in time, 
but the virial velocity of the halo decreases in general. The bottom row  shows that at z=2, the dynamic range in specific star formation rate across the $V_{vir}$ versus $M_{BH}$ plane is much smaller than at $z=0$.   
It can be concluded that, without additional quenching mechanisms, a stronger dependence of the feedback on $V_\mathrm{vir}$ and/or a weaker dependence on $M_\mathrm{BH}$ (with the appropriate re-adjustment of $\kappa$) at $z=2$ might improve the peak separation discrepancy. Favouring $V_{\mathrm{vir}}$ over $M_{\mathrm{BH}}$ in equation \ref{equ:feedback1} would lead to stronger feedback at early times, where halo virial velocities are high and black hole masses are low compared to their present values.

\begin{figure}
\begin{center}
\leavevmode
\includegraphics[scale=0.35]{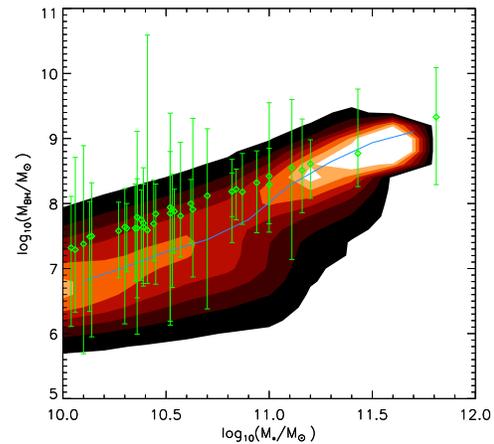}
\caption{The relation between galaxy central black hole mass and stellar mass at redshift $\sim$ 2. The filled contours represent the model predictions while the green data points are observational results corresponding to galaxies in a redshift interval of 1.5 $\le z \le$ 2.5.}
\label{fig:mbh_mstar}
\end{center}
\end{figure}

Imposing an ad hoc redshift dependence on the equations
that govern so-called radio- AGN feedback is certainly
not ideal.
Another possibility is to modify the black hole growth model. 
As can be seen from Figure \ref{fig:mbh_mstar}, the $z \sim 2$ quenched population 
of galaxies with stellar masses less than $\sim 10^{11} M_{\odot}$,
should arise from feedback by black holes with masses of $\sim 10^6 - 10^7 M_{\odot}$.
Black hole growth in the model is 
driven by galaxy mergers, particularly major mergers. There are however other viable
black hole feeding mechanisms, such as disc instabilities or deviations 
from axisymmetry (\cite{Bournaud11}), that could play a role, especially at higher 
redshifts, where star-forming galaxies are more gas-rich and so more likely to 
undergo such processes. In the Guo et al models, low mass bulges are produced by disk
instabilities, but black hole growth does not occur during this process.
The black hole mass vs. stellar mass relation predicted by 
the model, for $z \sim$ 2, is plotted in Figure \ref{fig:mbh_mstar} together with 
observational data taken from \cite{Simmons12}. At low stellar masses,
black holes are  slightly undermassive by about  $\sim$ 0.5 dex on average compared to  observations, though towards the high mass end the agreement is good.
In addition, as seen from the countours in Figure \ref{fig:mbh_mstar}, at all stellar masses less than
$2 \times 10^{11} M_{\odot}$, the models predict a tail of  black holes
that scatter to masses  more than a factor of 10 below the mean. This is not seen in the data.  
A secular black hole growth mechanism associated with disk instabilities  may help boost the average mass of this
population, resulting in more efficient quenching.

In addition, quenching mechanisms other than radio-AGN feedback may be required.    
In the local Universe, the frequency of large-scale FRI-type radio jets in galaxies with low mass black holes
is extremely low (\cite{Best05}). Almost all active galactic nuclei with black holes
of this mass are Seyfert galaxies or quasars, where much of the energy liberated during black
hole accretion emerges at ultraviolet or optical wavelengths.
It may thus be necessary to consider so-called `quasar mode' feedback. 
In the models, the quasar accretion mode occurs during gas-rich mergers, when progenitor black holes form through the
rapid accretion of cold gas. There is  
a starburst resulting from the merger, which is visible is the second panel of Figure 5 as a band of galaxies with high specific
star formation rates and black hole masses, and the merger remnant forms a galactic bulge.  
The merger consumes most of the gas, and supernovae eject gas from the galaxy and from the halo.  
In the current models, feedback from this  black hole accretion mode is not explicitly modelled. If feedback from
black hole formation is violent enough
to shut down star formation in a galaxy entirely for a few Gyr, 
this may result in more realistic SFRs in lower masses galaxies at $z\sim 2$. 
Indeed, examination of the top panel of  Figure \ref{fig:mbh_mstar} shows that models and data move into better agreement 
if galaxies with  bulge mass-to-total mass ratios greater than $\sim 0.5$ have been  
able to eject all their gas.

 \cite{Dubois12}, for example, have implemented a sub-grid model of black hole 
growth and feedback through quasar and radio mode in hydrodynamical 
cosmological simulations. They have shown that quasars are increasingly 
important at high redshift, where cold gas accretion triggers the quasar-mode 
of feedback. Gradually the cold gas reservoir is depleted by star formation, 
shock-heating of massive structures and feedback. Thus, the radio-mode becomes dominant at later times.
These authors did not, however, investigate whether their simulations could reproduce the observed evolution
of the bi-modal galaxy population. This is an important test of all galaxy modelling efforts that
include feedback from supermassive black holes. 

\bibliography{references}
\bibliographystyle{natbib}

\end{document}